\documentclass[english,aps, reprint, prd, twocolumn,floatfix,longbibliography]{revtex4-1}

\usepackage[T1]{fontenc}   
\usepackage{rotating}
\usepackage{booktabs}     
\usepackage{graphicx}   
\usepackage{float}
\usepackage{amsmath}
\usepackage{textcomp}
\usepackage{gensymb}
\usepackage{amssymb}
\usepackage{amstext}
\usepackage[colorinlistoftodos]{todonotes}
\usepackage{siunitx}
\usepackage{commath}
\usepackage{hyperref}

\usepackage[utf8]{inputenc}
\newcommand{\Refs}[1]{Refs.~\onlinecite{#1}}
\usepackage{babel} 

\begin{document}

\title{
Understanding the role of intramolecular ion-pair interactions in
conformational stability using an {\it ab initio} thermodynamic cycle 
}

\date{\today}     

\author{Sabyasachi Chakraborty $^1$}

\author{Kalyaneswar Mandal $^1$}

\author{Raghunathan Ramakrishnan $^1$}
\email{ramakrishnan@tifrh.res.in}

\affiliation{$^1$ Tata Institute of Fundamental Research Hyderabad, Hyderabad 500046, India}

\begin{abstract}
Intramolecular ion-pair interactions yield shape and functionality to many molecules.
With proper orientation, these interactions overcome
steric factors and are responsible for the compact structures of 
several peptides.
In this study, we present a thermodynamic cycle based on isoelectronic and alchemical mutation to estimate intramolecular ion-pair interaction energy. 
We determine these energies for 26 benchmark molecules with common
ion-pair combinations and compare them with results obtained using intramolecular symmetry-adapted perturbation theory. 
For systems with long linkers, the ion-pair energies evaluated 
using both approaches deviate by less than 2.5\% in vacuum phase. 
The thermodynamic cycle based on density functional theory
facilitates calculations of salt-bridge interactions in
model tripeptides with continuum/microsolvation modeling, and four large peptides: 
1EJG (crambin), 
1BDK (bradykinin), 
1L2Y (a mini-protein with a tryptophan cage), 
and 1SCO (a toxin from the scorpion venom). 
\end{abstract}

\maketitle   
\section{Introduction}
Ion-pair interactions are ubiquitous 
and are the fifth most prevalent in biological systems as salt-bridges\cite{de2017systematic}. There are many definitions of salt-bridge based on 
the constitution and spatial orientation of the interacting moieties\cite{kumar2002close,gilli2007pa,gilli2009predicting,gilli2010hydrogen,donald2011salt}.
In peptides, such interactions are predominantly found in guanidinium-carboxylate or ammonium-carboxylate ion pairs\cite{anslyn2006modern}.
Presence of only a few salt-bridges can have a significant impact on both the structure and the function of bio-macromolecules\cite{king2016formation,rivera2017one,pacheco2017intramolecular,smith2017enhancing,heiles2018competition,dianati2017rational,gorvin2018calcium,zhang2018salt,ben2020structure,mcmanus2019salt,zhao2016highly,hentzen2020lateral}. 
The stability of a salt-bridge depends on its local environment and exposure to polar solvents. 
Depending on these factors, they can be 
stabilising\cite{spek1998surface,vetriani1998protein,xiao1999electrostatic,kumar2000factors,kumar2000electrostatic,kumar2001thermophilic},
insignificant\cite{horovitz1990strength,serrano1990estimating,sun1991contributions,bycroft1991surface}, or even
destabilising\cite{hendsch1994salt}.
Experimentally, the presence of a salt-bridge can be inferred through 
nuclear magnetic resonance spectroscopy\cite{strop2000contribution} or infrared vibrational spectroscopy\cite{huerta2015structure,lotze2015structure}. 
While vibrational spectroscopy also
detects different salt-bridging patterns (such as end-on/side-on or monodentate/polydentate\cite{huerta2015structure}),
they cannot provide any insight on their relative thermodynamic stabilities.
The absolute value of the salt-bridge interaction energy of a folded protein is not
an experimental observable.
Between a folded and an unfolded state of a macromolecule, 
the Gibbs free energy of folding, $\Delta G_{\rm folding}$, 
can be experimentally determined in terms of stability constants. 
This quantity is essentially the sum of
contributions due to structural relaxation or strain, $\Delta G_{\rm strain}$,
salt-bridge (or ion-pair) interaction, $\Delta G_{\rm ip}$, and 
reorganization due to interactions with solvent molecules.
The contribution due to salt-bridge, $\Delta G_{\rm ip}$, can compensate for
the other two terms resulting in a net negative
$\Delta G_{\rm folding}$. 
Experimentally, $\Delta G_{\rm ip}$ is determined
using denaturation titrations of the wild-type protein and its single and double mutants\cite{ge2008salt,strop2000contribution}.
These experiments have been extremely useful to shed light on the strength of different salt-bridges across many different proteins.
The two salt-bridges (E6/R92 and E62/K46) in thermophilic ribosomal protein L30e stabilize the system by 0.5--1.2 kcal/mol\cite{chan2011stabilizing}.
A double mutant cycle analysis revealed the K11/E34 surface salt-bridge in ubiquitin to have a strength of 0.24 kcal/mol\cite{makhatadze2003contribution}, while the conserved salt-bridges (R69/D93 and R83/D75) in barnase revealed an interaction energy of 3.0--3.5 kcal/mol\cite{tissot1996importance}.
White \textit{et al.} used molecular dynamics simulations to investigate charged amino-acids in water, when interaction strength ($\Delta G_{\mathrm{int}}$) was found to vary from 0.5--2.0 kcal/mol\cite{white2013free}.
Typically, the stabilization due to salt-bridges in most proteins are within the 4--5 kcal/mol range\cite{anslyn2006modern}.
However, studies have shown that this value can be influenced by interactions with osmolytes\cite{tiwari2021interaction}.
In general, engineered surface salt-bridges are destabilizing but conserved salt-bridges present inside the protein are stabilizing\cite{takano2000contribution}.

We note that in denaturation experiments the interacting moieties in ion-pairs are in two very different paradigms: 
In the folded state, the ion-pair is not exposed to any solvent molecules, while in the unfolded state, it is exposed to both solvent molecules and ions.
Furthermore, when ion-pairs are exposed to the solvent media, the interaction strength is significantly quenched due to the dielectric constant of the solvent.
Perhaps, the ideal scenario of understanding unfolding must be one where contributions due to solvent can be completely ignored, but at this stage it is experimentally unfeasible.
Again, in double mutant experiments, the single mutants involve replacing one of the charged residues with a neutral one. 
The assumption is that the effects due to mutation will exactly cancels out.
However, the wild-type and mutated peptides may exhibit different conformational preferences resulting in non-cancellation of energy terms due to structural relaxation.
Another computational protocol would be to access a conformer of the biomolecule where the interacting groups are well-separated\cite{sapse2002role}. 
Not only does this exercise become increasingly challenging when it comes to estimating salt-bridges in large peptides, but more fundamentally, the strength of the salt-bridge is now dependent on the type of elongated conformer accessed.

Focusing only on the enthalpic contributions, a
theoretical estimation of intramolecular ion-pair
interaction energy, $E_{\rm ip}^{\rm intra}$, is more challenging than that of the 
intermolecular counterpart, $E_{\rm ip}^{\rm inter}$.
This is because the latter is a well-defined quantity related to the dissociation energy, while such analogy is 
not suitable for intramolecular bonding.
$E_{\rm ip}^{\rm inter}$ can be estimated using the supermolecule approach  
as the difference between the total energies of the complex and its isolated monomers.
$E_{\rm ip}^{\rm inter}$ determined for model salt-bridges provides insights into 
$E_{\rm ip}^{\rm intra}$ of proteins. 

Besides the supermolecule approach, intermolecular interaction energies can also 
be estimated using 
energy decomposition analysis (EDA)\cite{ziegler1977calculation,morokuma1971molecular,su2009energy,hopffgarten2012energy,zhao2018energy} 
or symmetry-adapted perturbation theory (SAPT)\cite{jeziorski1994perturbation,szalewicz2012symmetry,hohenstein2012wavefunction,gonthier2014quantification,corminboeuf2014quantifying,patkowski2020recent,parrish2015communication,pastorczak2015intramolecular,hesselmann2002intermolecular,hesselmann2005density,parker2014levels}. Both EDA and SAPT
provide a breakdown of interaction energies quantifying various physical effects. 
These methods have been benchmarked using results from the 
supermolecule approach based on 
highly accurate wavefunction methods as references. 
The accuracy of SAPT is limited by the
reference wavefunction used to model the interacting fragments and the order of 
perturbation corrections. 
In the intramolecular
analog of SAPT (I-SAPT), Hartree--Fock (HF) molecular orbitals (MOs) are localized on the interacting fragments and the linker. 
Collectively, this procedure is denoted as F/ISAPT0, where 0 indicates the use of HF states for the fragments prior to the application of perturbation theory\cite{parrish2014quantum}. 
The initial applications of F/ISAPT0 to organic molecules such as pentanediol have yielded physically
meaningful partitioning of the interaction energy, 
of $E_{\rm ip}^{\rm intra}$\cite{parrish2015communication}.
In general, the SAPT formalism delivers excellent predictions for weak non-covalent interactions
in charge-neutral systems, while they under-perform for ionic systems\cite{lao2012breakdown,li2014quantum,lao2015accurate}. 
Hence, for zwitterions 
stabilized by intramolecular ion-pair
interactions, it is desirable to compare F/ISAPT0's predictions to results from other approaches.

In this study, we present a thermodynamic cycle to estimate $E_{\rm ip}^{\rm intra}$ by introducing isoelectronic
atomic mutations. 
For model dimers with intermolecular ionic bonding, we benchmark SAPT0 with two basis sets against CCSD(T) references.
We use the best performing basis set in combination with F/ISAPT0 method, and determine intramolecular ion-pair
interactions for a set of 26 molecules. These results are compared with $E_{\rm ip}^{\rm intra}$ predictions using
the thermodynamic cycle based on total energies from
the domain-based local pair natural orbital coupled cluster with singles doubles and perturbative triples method, DLPNO-CCSD(T)\cite{riplinger2013efficient,riplinger2013natural} and various density funcitonal theory (DFT) approximations. 
We probe the convergence of $E_{\rm ip}^{\rm intra}$ with different fragmentation procedures
in vacuum and implicit solvation regimes.
We explore the effect of microsolvation on structural preferences, salt-bridge interaction energies, and conformer stabilities of four model tripeptides. 
Finally, we estimate salt-bridge interaction energies in four large peptides and rationalize their experimentally noted geometries.

\section{Theory}\label{Theory}
\begin{figure*}[!htbp]
    \centering
    \includegraphics[width=\linewidth]{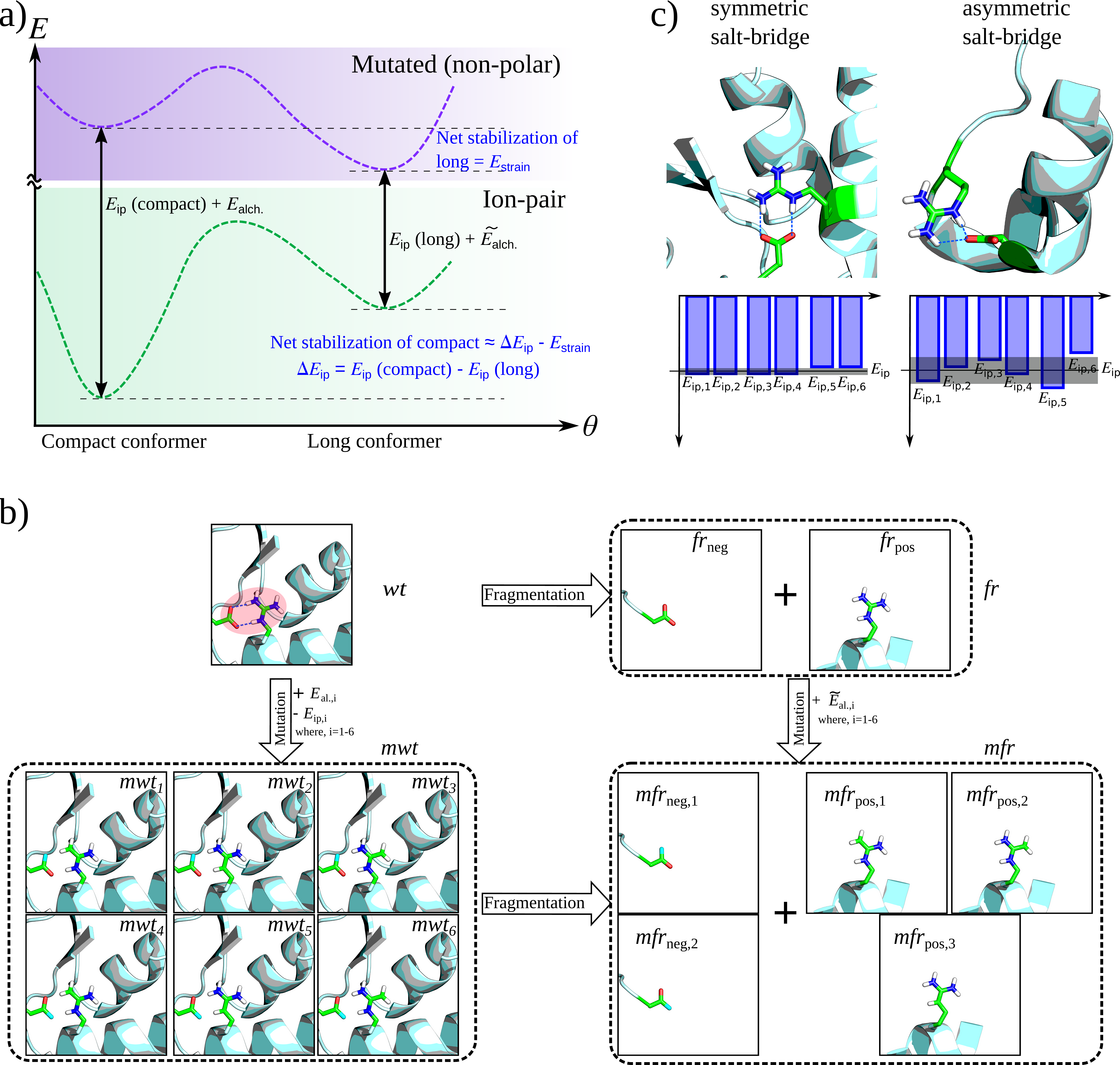}
    \caption{
    Development of the thermodynamic scheme. 
    a) Schematic illustration of typical potential energy profiles of a molecule in the zwitterionic form (bottom) and 
    in its alchemically mutated non-polar form (top). 
    The horizontal axis is denoted by $\theta$, a general reaction coordinate. 
    For simplicity, effect of hydration is not considered. 
    Net stabilization of the compact structure of an ion-pair can be estimated by assuming a cancellation of 
    alchemical changes due to introduction of charge neutral atoms or groups, $E_{\rm alch.}\thickapprox\widetilde{E}_{\rm alch.}$.
    b) Thermodynamic cycle for estimating intramolecular ion-pair interaction energy, $E_{\rm ip}$.
    As an example, we consider the salt-bridging interaction between the guanidinium and carboxylate moieties (highlighted with a red ellipse). Here, \textit{wt} denotes the wild-type variant of the molecule.
    Atomistic mutations to remove charges on moieties leads to mutated wild-type variants denoted by \textit{mwt}$_{\rm i}$, where i runs from 1 to 6, spanning all possible mutation paths.
    We now fragment \textit{wt} to two separate units, one containing the positively charged moiety and the other containing the negatively charged moiety, denoted as {\it fr}$_{\rm pos}$ and {\it fr}$_{\rm neg}$ respectively.
    Dangling bonds in both moieties are saturated. 
    Thus, mutating {\it fr}$_{\rm pos}$, and {\it fr}$_{\rm neg}$ leads to {\it mfr}$_{\rm pos,j}$, and
    {\it mfr}$_{\rm neg,k}$ where $j$ 
    covers two mutation possibilities in the carboxylate,  while $k$ covers the three possibilities in guanidinium.
    Cartesian coordinates are preserved during the two mutation processes, namely, {\it wt}$\longrightarrow$\textit{mwt}$_i$ and {\it fr}$_{\rm pos/neg}$ $\longrightarrow${\it mfr}$_{\rm pos,j/neg,k}$.
    c) Distribution of $E_{{\rm ip},i}$ for symmetric and asymmetric salt-bridges, $E_{\rm ip}^{\rm intra}$ is taken as the mean. 
    }
    \label{fig01}
\end{figure*}

Here, we focus on the ion-pair interaction energy, $E_{\rm ip}$, and its contribution to the stability of conformers. 
An idealized potential energy profile of an ion-pair system connecting a compact conformer and a longer one is presented 
in FIG.\ref{fig01}a.
The net ion-pair bonding in the compact conformer, $\Delta E_{\rm ip}$, overcomes the
steric strain associated with folding. 

Hence, on the potential energy surface (PES) of the non-polar mutant, the longer conformer is energetically favored.
In principle, it is possible to apply the schematics shown in FIG.\ref{fig01}a to estimate the difference in
$E_{\rm ip}$ between a compact and a long conformer ($\Delta E_{\rm ip}$). 
In practice, the interest lies in estimating $E_{\rm ip}$ for an 
experimentally observed folded geometry such as those collected 
in the protein data bank (PDB)\cite{berman2003announcing},
without depending on a reference unfolded geometry. 
While determining an unfolded conformer is possible by exploring the torsional PES 
(as routinely explored in molecular dynamics studies), our thermodynamic scheme discussed below, avoids this step.

In FIG.\ref{fig01}b we present a scheme to 
estimate $E_{\rm ip}^{\rm intra}$ of an ion-pair
system using a thermodynamic cycle.
The specific case of ammonium-carboxylate ion-pair is used as an example. 
Here, {\it wt} denotes the system with an intramolecular ion-pair bonding, and {\it mwt}$_i$ is obtained by 
combinatorially mutating the polar groups with isoelectronic substitutions.
Thus, N atoms in the positively charged polar moiety in {\it wt} are substituted with C atoms while 
in the negatively charged moiety, O atoms are replaced by F.
The difference between
the total energies of $wt$ and $mwt$ quantifies the ion-pair interaction that is 
present in $wt$ while absent in $mwt$,
and contributions due to the net change in the external potential 
and nuclear repulsion
going from $wt$ to $mwt$:
\begin{eqnarray}
   E_{ wt} - E_{{ mwt, i}} = E_{{\rm ip,} i} + E_{{ \rm al.}}
   \label{eq01}
\end{eqnarray}
The alchemical contribution, $E_{\rm al.}$, can be estimated separately by using the total energies of the fragmented monomers of
$wt$ and $mwt$, where the Cartesian coordinates of atoms of the polar groups are kept frozen. 
Hence,
\begin{eqnarray}
   E_{fr} - E_{{mfr},i} = \widetilde{E}_{{\rm al.}}
   \label{eq02}
\end{eqnarray}
Fragmenting \textit{wt} to infinitely separated units comprising the charged moieties and a limited degree of their local chemical environment yields two fragmented systems, denoted as \textit{fr}$_{\rm pos}$, and \textit{fr}$_{\rm neg}$ (commonly denoted as \textit{fr}).
These polar moieties are saturated with terminal groups of varying length whose performances are later investigated.
Further, for all possible mutation paths in the wild type system, \textit{wt}$\longrightarrow$\textit{mwt}$_{i}$, 
there are corresponding mutations in the fragments, \textit{fr}$_{\rm pos/neg}$ $\longrightarrow$\textit{mfr}$_{{\rm pos},j/{\rm neg},k}$.
Our scheme can be applied to determine $E_{\rm ip}$ with microsolvation to capture solvent effects. 
In this regime, we hydrate our target molecules in a polarizable continuum medium and the geometries are optimized to capture
reorganization energies. 
When a molecule with intramolecular ion-pair interactions is relaxed with a fixed number of water molecules, the resulting fragments will also contain the same number of water molecules with same orientation as in {\it wt/mwt}.

Combining (\ref{eq01}) and (\ref{eq02}) we arrive at
\begin{eqnarray}
    E_{{\rm ip},i} \thickapprox \left( E_{wt} - E_{{mwt},i}  \right) - \left( E_{fr} - E_{{mfr},i}  \right) 
   \label{eq03}
\end{eqnarray}
for the $i$-th mutation path.
The approximation holds as long as $E_{{\rm al.}}\thickapprox \widetilde{E}_{{\rm al.}}$ which can be expected when the 
interacting moieties are appropriately selected 
to adequately capture polarization effects on the linker atoms.
The final value of $E_{\rm ip}^{\rm intra}$ for a given geometry of {\it wt}
is estimated by averaging over all possible alchemical mutations, 
$E_{\rm ip}^{\rm intra}=\langle E_{{\rm ip},i} \rangle$. 

The standard deviation
with respect to the mean indicates asymmetry of the ion-pair bonding pattern as shown
schematically in FIG.\ref{fig01}c. 
Hence, an asymmetric bonding pattern will result in distinct $E_{\rm ip}$ 
values whose standard deviation corresponds to the method's uncertainty.
For symmetric bonding patterns, the standard deviation vanishes as all mutation paths are equivalent.

\section{Computational Methods}\label{Sec:Computational_Methods}
For four large peptides, 1EJG, 1BDK, 1L2Y, and 1SCO, we collected geometries from PDB. 
For all other systems,
we performed geometry optimizations using the $\omega$B97X-D3-DFT\cite{lin2013long,grimme2010consistent}
method with the def2-TZVP basis set. 
This method was used because of its ability to converge to
zwitterionic molecules while preserving the intended bond connectivities\cite{senthil2021troubleshooting}.
Solvent effects (water medium) were modeled using the conductor-like polarizable continuum medium (CPCM) approach\cite{barone1998quantum,york1999smooth}. 
Explicit solvent interactions were modeled 
using the microsolvation approach where 2, 4, or 8 explicit water molecules were included in the DFT/CPCM
geometry optimizations. 
In microsolvation investigations, different conformers were generated starting with randomly sampled arrangements of the water molecules. 
For a given solute, the center of the centroid--centroid distance of the polar moieties involved in non-covalent ion-pair interaction was chosen as a center of a sphere on whose surface the water molecules were randomly dispersed.
This implies that in the initial geometries generated, the water molecules were close to the ion-pair containing end of the molecules.
Both solutes and solvents were then allowed to undergo geometry relaxation.
In vacuum phase geometry optimizations, N-H bond lengths were constrained at the cationic
terminals in order to converge to an ion-pair local minimum and prevent converging to a  
thermodynamically more stable proton-transferred structure. 
Such constraints were not required with CPCM and microsolvation for most of the
systems studied here. All single-point calculations were performed on the $\omega$B97X-D3/def2TZVP geometries. For the
aforestated large peptides, such calculations were done using the PDB geometries. 
DLPNO-CCSD(T)\cite{riplinger2013efficient,riplinger2013natural} 
calculations were performed with the aug-cc-pVTZ basis set and \texttt{TightPNO} settings.
DFT and DLPNO-CCSD(T) calculations were performed using 
ORCA 5.0\cite{neese2012orca,neese2018software} with the resolution-of-identity (RI) approximation\cite{vahtras1993integral,kendall1997impact} for Coulomb (J) and `chain-of-spheres' (COS) algorithm for exchange integrals (RIJCOSX). For RI calculations we used the Weigend auxiliary basis sets\cite{weigend2006accurate}. 
DFT calculations were done with the default grid, \texttt{defgrid2}.
Intramolecular ion-pair interaction energies were modeled with 
F/ISAPT0\cite{parrish2015communication} using aug-cc-pVTZ basis set. This basis set was selected after comparing
its accuracy with that of jun-cc-pVDZ for modeling SAPT0-level intermolecular molecular 
ion-pair interaction energies. All SAPT calculations were performed with the code Psi4\cite{parrish2017psi4}.

The basis set effect on ion-pair interaction energies was benchmarked with $\omega$B97X-D3 energies calculated
using def2-SVP, def2-TZVP, def2-SVPD, and def2-TZVPD basis sets.
We have also benchmarked the 
performance of 10 different DFT approximations from different rungs of \textit{Jacob's ladder} using
the def2-SVPD basis set. For every functional benchmarked, the recommended semi-empirical dispersion corrections were included. Hence, we included Grimme's D3 correction with Becke-Johnson damping (D3BJ)\cite{grimme2011effect,becke2005density,johnson2005post,johnson2006post} or D3Zero\cite{grimme2010consistent} for functionals without long-range corrections. 
For PBE\cite{perdew1996phys}, BLYP\cite{becke1988density}, B3LYP\cite{becke1993becke}, PBE0\cite{adamo1999toward}, and TPSS\cite{tao2003climbing} we included D3BJ, while for M06-2X\cite{zhao2008m06} we invoked D3Zero. 
For functionals with implicit long-range corrections such as LC-PBE\cite{iikura2001long,najibi2021analysis}, LC-BLYP\cite{tawada2004long}, CAM-B3LYP\cite{yanai2004new}, and $\omega$B97X-D3\cite{lin2013long} no additional dispersion corrections were included. Intramolecular ion-pair interaction energies of peptides reported here are at the $\omega$B97X-D3/def2-SVPD level.

\section{Results and Discussion}
In FIG.~\ref{fig01}, we discussed the influence of an ion-pair interaction on the net stability of the compact conformer over the long conformer. 
While the compact conformer is most stable when its polar groups are suitably oriented for maximal interaction, this comes at the cost of internal strain. 
In large molecules, this balance between strain and ion-pair bonding along with myriad other non-covalent interactions determine the global minimum's structure\cite{de2017systematic}. 
Although studies have quantified the effect of interactions between polar groups from a super-molecule approach\cite{white2013free,kurczab2018salt}, or as relative free energy changes using molecular dynamics simulations\cite{ahmed2018well,huang2018generalized}
it is of interest to directly estimate such intramolecular interaction energies for a given structure. 

In this study, we pursue this goal with F/ISAPT0 and a thermodynamic cycle. 
To benchmark these methods, we design two sets of model systems. 
The first set comprises nine dimers stabilized by intermolecular ion-pair bonding. The interaction energies of these
molecules were estimated in a supermolecule fashion with the CCSD(T)/aug-cc-pVTZ method. Using these values
as reference, we benchmarked SAPT0/jun-cc-pVDZ and SAPT0/aug-cc-pVTZ methods, and found the aug-cc-pVTZ to be
more suitable for SAPT modeling of ion-pair systems. Subsequently, we modeled intramolecular ion-pair interactions 
in a second set with 26 molecules using the
aug-cc-pVTZ basis set in combination with the F/ISAPT0 method. For this set, we also
benchmark $E_{\rm ip}$(Eq.~\ref{eq03}) determined with the thermodynamic cycle based on different DFT methods.
With the best set up, we report salt-bridge interaction energies of model tripeptides and large proteins in solvent medium.
\begin{figure}[!hbtp]
    \centering
    \includegraphics[width=\linewidth]{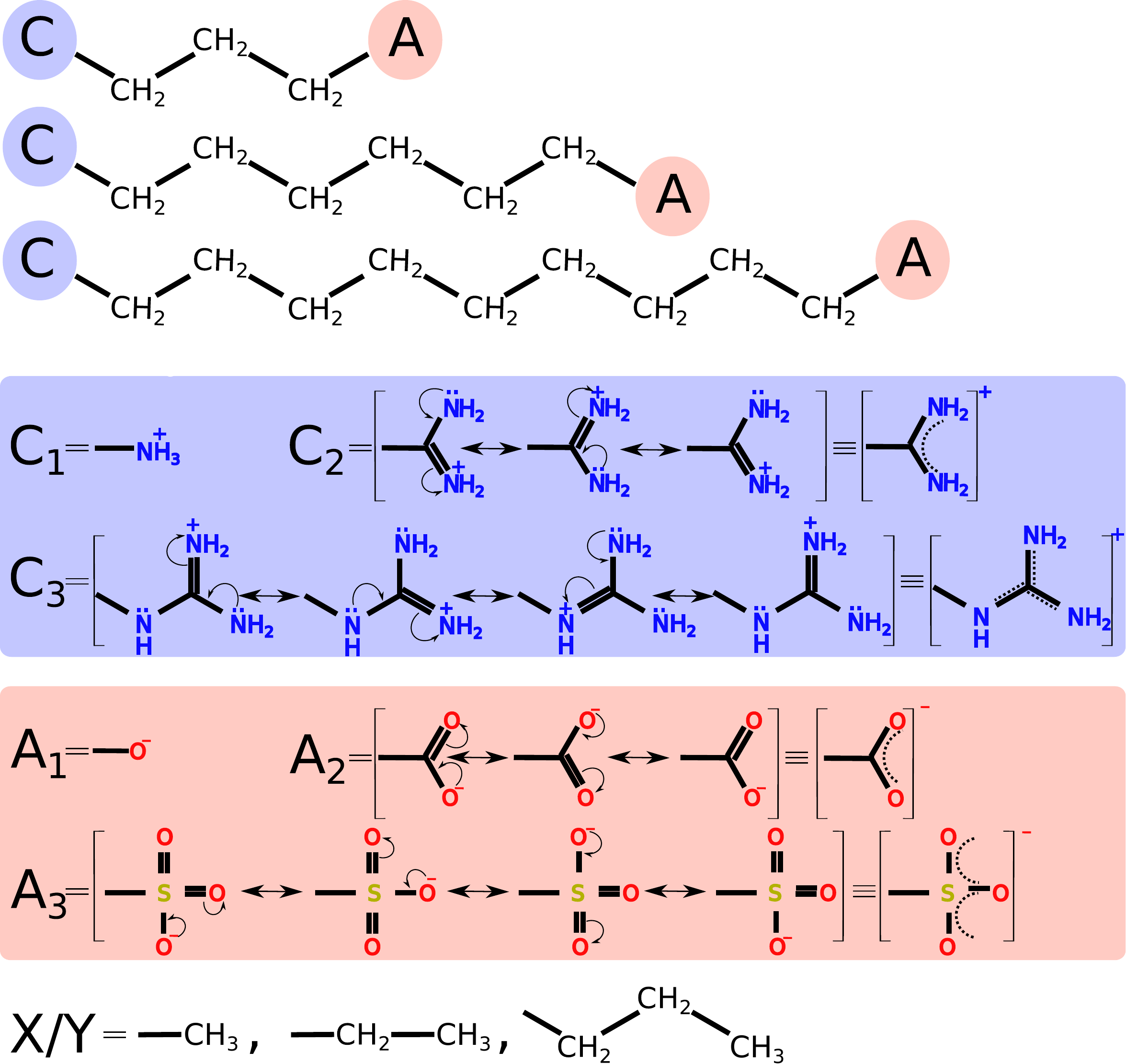}
    \caption{
    Constitution of benchmark molecules with intramolecular ion-pair bonding. 
    The set is formed by combining 3 cationic terminals---C$_1$: monodentate (ammonium),
    C$_2$: bidentate (formamidinium), C$_3$: tridentate (guanidinium) with 
    3 anionic terminals---A$_1$: monodentate (hydroxy), A$_2$: bidentate (carboxylate), 
    A$_3$: tridentate (sulphate) via three linkers---L$_1$: -(CH$_2$)$_3$-,
    L$_2$: -(CH$_2$)$_6$-, L$_3$: -(CH$_2$)$_9$-.
    The various terminal groups used when fragmenting the system into monomers are denoted by X/Y: -CH$_3$, -CH$_2$CH$_3$,\& -CH$_2$CH$_2$CH$_3$.
    }
    \label{fig04}
\end{figure}
\subsection{Model ion-pair systems}
In FIG.~\ref{fig04} we describe the composition of model ion-pair systems considered in this study.
The polar groups comprises three cationic (C$_n$), three anionic (A$_n$) moieties that are biochemically
relevant\cite{kurczab2018salt}. 
The diverse resonance structures of these polar moieties facilitate various bonding patterns 
as seen, for example, in guanidinium--carboxylate salt-bridges\cite{huerta2015structure}. 
The cation set includes ammonium (C$_1$), formamidinium (C$_2$), and guanidinium (C$_3$) moieties with 1, 2 and 3 resonance structures, respectively, while the anion set includes oxide (A$_1$), carboxylate (A$_2$), and sulphate (A$_3$) moieties with a similarly increasing number of resonance structures. 

The fist set of model systems used for quantifying $E_{\rm ip}^{\rm inter}$ were formed by attaching methyl groups to the
polar moieties, directed away from the ion-pair bonding. This set contains 9 dimers formed by combining 3 cations with 3 anions. For
each system, we estimated $E_{\rm ip}^{\rm inter}$ for 3 different centroid-centroid distances, resulting in 27 energies. 
The second set of  model systems to quantify $E_{\rm ip}^{\rm intra}$ comprise these same polar moieties by covalently connected
via three different linker (L$_n$) fragments, $n = 1, 2, 3$.
For linkers, we chose linear alkane chains with 3 (L$_1$), 6 (L$_2$), and 9 (L$_3$) methylene units. 
These non-polar linkers are not expected to interfere with the salt-bridging interactions. 
Of the 27 possibilities, C$_2$-L$_1$-A$_1$ was ignored because during geometry optimization it cyclized to an amide.

\subsection{Basis set for SAPT modeling of ion-pair interactions}
In Table.~\ref{tab:basis_set_benchmark} we report SAPT0-level $E_{\rm ip}^{\rm inter}$ using jun-cc-pVDZ and aug-cc-pVTZ basis sets
for 9 model ion pairs with three different centroid-centroid distances.
In these complexes, the polar moieties faced each other while the methyl groups were pointed outwards. 
Further, for C$_1$--A$_{1/2/3}$ and C$_{2/3}$--A$_1$ type complexes, we accessed the 3.0 \AA, 5.0 \AA, and 7.0 \AA{} centroid--centroid distances, while the remaining complexes were placed at 4.0 \AA, 5.0 \AA, and 7.0 \AA{} to avoid small interatomic distances.
We compared the SAPT0 results with those from a supermolecule scheme based on CCSD(T)/aug-cc-pVTZ energies.
For short centroid--centroid distances, the interaction energies are high.
On an average, we note that the performance of SAPT0/jun-cc-pVDZ to be inferior to SAPT0/aug-cc-pVTZ, with the largest deviations noted 
at short contacts.
Hence, in the remainder of the study, we use the aug-cc-pVTZ basis set for modeling $E_{\rm ip}^{\rm intra}$ at the F/ISAPT0 level.

\begin{table}[!htbp] 
    \centering
    \caption{
    Deviation of intermolecular ion-pair interaction energies of 9 dimer model systems 
    estimated using SAPT0/aug-cc-pVTZ and 
    SAPT0/jun-cc-pVDZ from CCSD(T)/aug-cc-pVTZ reference. The mean absolute deviation of both methods are reported in the last row.
    Values in parenthesis in the first column the centroid--centroid distances between the monomers in \AA.
    All other values are reported in kcal/mol. 
    }
    \resizebox{\linewidth}{!}{
    \begin{tabular}{l r r r }
    \hline
     & &\multicolumn{2}{l}{SAPT0}\\
     \cline{3-4}
     System & \multicolumn{1}{l}{Reference} &  \multicolumn{1}{l}{aug-cc-pVTZ}& \multicolumn{1}{l}{jun-cc-pVDZ} \\
    \hline
 C$_1$ -- A$_1$ (3.0) &  -103.25  &   3.94  &     3.63 \\          
 C$_1$ -- A$_1$ (5.0) &   -60.09  &   4.15  &     4.61 \\          
 C$_1$ -- A$_1$ (7.0) &   -60.09  & -14.16  &   -13.84 \\          
 
 C$_1$ -- A$_2$ (3.0) &  -121.52  &   1.54  &    -1.29 \\          
 C$_1$ -- A$_2$ (5.0) &   -73.24  &  -0.06  &     0.21 \\          
 C$_1$ -- A$_2$ (7.0) &   -50.34  &   0.25  &     0.54 \\          
 
 C$_1$ -- A$_3$ (3.0) &  -110.43  &  -1.02  &    -4.21 \\          
 C$_1$ -- A$_3$ (5.0) &   -68.60  &  -0.20  &    -0.18 \\          
 C$_1$ -- A$_3$ (7.0) &   -48.63  &   0.14  &     0.35 \\          
 
 C$_2$ -- A$_1$ (3.0) &  -131.13  &   0.88  &    -2.55 \\          
 C$_2$ -- A$_1$ (5.0) &   -84.76  &  -0.75  &    -0.69 \\          
 C$_2$ -- A$_1$ (7.0) &   -49.49  &  -0.07  &     0.22 \\          
 
 C$_2$ -- A$_2$ (4.0) &  -122.10  &   0.74  &    -2.37 \\          
 C$_2$ -- A$_2$ (5.0) &   -89.67  &  -1.01  &    -1.30 \\          
 C$_2$ -- A$_2$ (7.0) &   -55.88  &  -0.10  &     0.13 \\          
 
 C$_2$ -- A$_3$ (4.0) &  -103.93  &  -2.39  &    -5.53 \\          
 C$_2$ -- A$_3$ (5.0) &   -80.47  &  -1.44  &    -2.13 \\          
 C$_2$ -- A$_3$ (7.0) &   -53.36  &  -0.21  &    -0.10 \\          
 
 C$_3$ -- A$_1$ (3.0) &  -126.03  &   2.61  &    -1.22 \\          
 C$_3$ -- A$_1$ (5.0) &   -71.83  &  -0.15  &     0.04 \\          
 C$_3$ -- A$_1$ (7.0) &   -48.06  &   0.10  &     0.34 \\          
 
 C$_3$ -- A$_2$ (4.0) &  -106.55  &   0.82  &    -2.18 \\          
 C$_3$ -- A$_2$ (5.0) &   -82.19  &  -0.15  &    -0.58 \\          
 C$_3$ -- A$_2$ (7.0) &   -53.01  &   0.29  &     0.48 \\          
 
 C$_3$ -- A$_3$ (4.0) &   -99.95  &  -0.92  &    -4.61 \\          
 C$_3$ -- A$_3$ (5.0) &   -77.64  &  -0.95  &    -1.64 \\          
 C$_3$ -- A$_3$ (7.0) &   -51.40  &   0.11  &     0.22 \\
    \hline
                &  & 1.08 & 1.70 \\
    \hline
    \end{tabular}
    }
    \label{tab:basis_set_benchmark}
\end{table}

\subsection{Intramolecular ion-pair interactions in model systems}
\begin{table*}[!htbp]
    \centering
    \caption{
    Comparison of ISAPT-based intramolecular interaction energies, $E^{\rm int}$, 
    between the polar groups for 26 benchmark systems (see FIG.\ref{fig04}) 
    with intramolecular
    ion-pair interaction energies, $E_{\rm ip}$, estimated using the thermodynamic
    cycle (Eq.\ref{eq03}) with DLPNO-CCSD(T) and DFT total energies. 
    ISAPT and DLPNO-CCSD(T) calculations were performed using the 
    aug-cc-pVTZ (AVTZ) basis set while $\omega$B97X-D3 DFT calculations were done
    with the def2- family of basis sets. 
    Mean absolute deviation (MAD), and root-mean-square deviation (RMSD) in kcal/mol along with
    mean percentage absolute deviation (MPAD) are reported. 
    Thermodynamic cycle based
    estimation was performed for terminal groups of varying length; these groups
    are denoted as X and Y in FIG.\ref{fig04}.
    }
 \begin{tabular}{l r r r l r r r  l r r r  l r r r }
 \hline
 \multicolumn{1}{l}{Methods} &
 \multicolumn{4}{l}{C$_x$L$_1$A$_y$-systems (8)}&
 \multicolumn{4}{l}{C$_x$L$_2$A$_y$-systems (9)}&
 \multicolumn{4}{l}{C$_x$L$_3$A$_y$-systems (9)}&
 \multicolumn{3}{l}{All systems (26)}\\
 \cline{2-4}\cline{6-8} \cline{10-12}\cline{14-16}
 \multicolumn{1}{l}{} &
 \multicolumn{1}{l}{MAD}& \multicolumn{1}{l}{RMSD}& \multicolumn{1}{l}{MPAD}&\multicolumn{1}{l}{} &
 \multicolumn{1}{l}{MAD}& \multicolumn{1}{l}{RMSD}& \multicolumn{1}{l}{MPAD}&\multicolumn{1}{l}{} &
 \multicolumn{1}{l}{MAD}& \multicolumn{1}{l}{RMSD}& \multicolumn{1}{l}{MPAD}&\multicolumn{1}{l}{} &
 \multicolumn{1}{l}{MAD}& \multicolumn{1}{l}{RMSD}& \multicolumn{1}{l}{MPAD} \\
 \hline 
\multicolumn{16}{l}{Terminal group: -CH$_3$} \\
$\omega$B97X-D3/def2-SVP    & 24.36 &11.09 &  21.96 && 20.04 &9.28 &  14.89 && 21.68 &8.99 &  15.44 && 21.94 &9.93 &  17.26 \\
$\omega$B97X-D3/def2-TZVP   & 18.16 &10.50 &  16.63 && 12.38 &7.34 &   9.16 && 13.36 &6.77 &   9.46 && 14.50 &8.63 &  11.56 \\
$\omega$B97X-D3/def2-SVPD   & 15.97 &10.73 &  14.72 &&  9.78 &6.81 &   7.25 && 10.19 &5.82 &   7.19 && 11.82 &8.40 &   9.53 \\
$\omega$B97X-D3/def2-TZVPD  & 16.07 &10.58 &  14.91 &&  9.75 &6.50 &   7.27 && 10.14 &5.39 &   7.19 && 11.83 &8.18 &   9.59 \\
DLPNO-CCSD(T)/AVTZ          & 16.08 &10.33 &  14.87 &&  9.52 &6.35 &   7.08 &&  9.74 &5.48 &   6.89 && 11.62 &8.11 &   9.41 \\
\\
\multicolumn{16}{l}{Terminal group: -CH$_2$CH$_3$} \\
$\omega$B97X-D3/def2-SVP    & 17.52 & 9.78 &  16.55 && 12.82 &5.47 &   9.80 && 14.53 &3.93 &  10.52 && 14.86 &6.98 &  12.13 \\
$\omega$B97X-D3/def2-TZVP   & 13.04 &10.12 &  12.58 &&  6.83 &4.92 &   5.29 &&  7.33 &3.02 &   5.32 &&  8.92 &7.05 &   7.54 \\
$\omega$B97X-D3/def2-SVPD   & 12.43 &10.55 &  11.89 &&  5.41 &5.19 &   4.21 &&  4.83 &3.25 &   3.52 &&  7.37 &7.40 &   6.33 \\
$\omega$B97X-D3/def2-TZVPD  & 12.47 &10.53 &  12.00 &&  5.34 &5.11 &   4.19 &&  4.91 &2.83 &   3.60 &&  7.38 &7.34 &   6.39 \\
DLPNO-CCSD(T)/AVTZ          & 12.34 &10.45 &  11.89 &&  4.88 &4.77 &   3.84 &&  4.31 &2.72 &   3.17 &&  6.98 &7.28 &   6.09 \\
\\
\multicolumn{16}{l}{Terminal group: -CH$_2$CH$_2$CH$_3$} \\
$\omega$B97X-D3/def2-SVP    & 13.09 &10.02 &  12.79 &&  9.48 &5.08 &   7.30 && 11.41 &2.91 &   8.29 && 11.26 &6.70 &   9.33 \\
$\omega$B97X-D3/def2-TZVP   & 11.13 &10.50 &  10.80 &&  4.97 &4.80 &   3.88 &&  4.83 &2.44 &   3.53 &&  6.81 &6.94 &   5.89 \\
$\omega$B97X-D3/def2-SVPD   & 10.71 &10.86 &  10.32 &&  4.57 &5.07 &   3.60 &&  3.25 &2.57 &   2.40 &&  6.00 &7.24 &   5.25 \\
$\omega$B97X-D3/def2-TZVPD  & 10.87 &10.88 &  10.47 &&  4.55 &5.05 &   3.59 &&  2.97 &2.52 &   2.22 &&  5.95 &7.26 &   5.23 \\
DLPNO-CCSD(T)/AVTZ          & 10.93 &10.87 &  10.51 &&  4.21 &4.76 &   3.32 &&  2.73 &2.49 &   2.07 &&  5.77 &7.25 &   5.10 \\
\hline
 \end{tabular}
 \label{tab:basis_benchmark_1}
\end{table*} 
For 26 model systems, we determine $E_{\rm ip}^{\rm intra}$ using the thermodynamic cycle (Eq.\ref{eq03}) and compare
the results with F/ISAPT0/aug-cc-pVTZ values ( $E^{\rm int}_{\rm ISAPT}$). 
The increasing size of the system with linker lengths rendered CCSD(T)/aug-cc-pVTZ
level calculations prohibitively expensive. Hence, we resorted to the DLPNO variant of CCSD(T) for the 
thermodynamic cycle, and pare. 
Since the inherent limitations of both methods stem from different sources, at the limit of their mutual convergence
their predictions can be expected to be closer to the reality.
In Table.~\ref{tab:basis_benchmark_1}, we report on the deviation 
in interaction energies ($E^{\rm int}_{\rm ISAPT}$) determined by F/ISAPT0 and
$E_{\rm ip}$ using Eq.~\ref{eq03} with total energies from
DLPNO-CCSD(T) and $\omega$B97X-D3-DFT levels 
for various choices of terminal groups: -CH$_3$, -CH$_2$CH$_3$, and -CH$_2$CH$_2$CH$_3$. 

In general, for all choices of terminal groups, and across all methods, the deviation between $E^{\rm int}_{\rm ISAPT}$ 
and $E_{\rm ip}$ [DLPNO-CCSD(T)] largely follows the order L$_1$ $>$ L$_2$ $>$ L$_3$. 
This trend is reflected better in the standard deviation (RMSD) and mean percentage absolute 
deviation (MPAD) than the mean absolute deviation (MAD). 
In L$_1$ systems, the net ion-pair interaction is not only mediated through space but also through polarization of the short linker.
This through-bond contribution is not accounted for in $E^{\rm int}_{\rm ISAPT}$.
However, in the case of longer linkers (L$_2$ and L$_3$), the through-bond contribution to $E^{\rm int}_{\rm ISAPT}$ 
becomes negligible justifying the MO localization scheme adopted. Hence, the F/ISAPT0 modeling  of ion-pair interaction is most appropriate when the polar groups are separated by sufficiently long linkers. 

\begin{table}[!htbp] 
    \centering
    \caption{
    Deviation of ion-pair interaction energies of 9 C$_x$L$_3$A$_y$ benchmark systems 
    estimated using F/ISAPT0/aug-cc-pVTZ and the thermodynamic
    cycle (Eq.\ref{eq03}) based on various density functional approximations with the def2-SVPD basis set. 
    For comparison, 
    DLPNO-CCSD(T)/aug-cc-pVTZ and $\omega$B97X-D3 values from Table.~\ref{tab:basis_benchmark_1} are also provided. 
    Mean absolute deviation (MAD), root-mean-square deviation (RMSD),
    mean percentage absolute deviation (MPAD) are reported in kcal/mol.  
    }
    \begin{tabular}{l r r r}
    \hline
     \multicolumn{1}{l}{Methods} &
     \multicolumn{1}{l}{MAD}& \multicolumn{1}{l}{RMSD}&\multicolumn{1}{l}{MPAD} \\
    \hline
PBE (D3BJ)       & 3.32 & 2.29 & 2.51  \\
BLYP (D3BJ)      & 2.78 & 2.39 & 2.11  \\
TPSS (D3BJ)      & 2.93 & 2.44 & 2.20  \\
LC-PBE           & 4.54 & 2.14 & 3.32  \\
LC-BLYP          & 3.24 & 2.23 & 2.43  \\
PBE0 (D3BJ)      & 3.34 & 2.33 & 2.47  \\
B3LYP (D3BJ)     & 3.01 & 2.65 & 2.26  \\
M06-2X (D3Zero)  & 3.49 & 3.18 & 2.59  \\
 CAM-B3LYP       & 3.01 & 2.63 & 2.25  \\
$\omega$B97X-D3  & 3.25 & 2.57 & 2.40  \\
DLPNO-CCSD(T)    & 2.73 & 2.49 & 2.07  \\
    \hline
    \end{tabular}
    \label{tab:DFT_benchmark_2}
\end{table}

In the thermodynamic cycle, we do not employ 
localized orbitals in order to describe the electron density as realistically as possible. Hence, when the spatial
extent of the polar group is confined only until the nearest neighbor group (terminal group: -CH$_3$) 
and further through-bond polarization is not captured. 
Therefore, increasing the length of the terminal group from -CH$_3$
to -CH$_2$CH$_2$CH$_3$ results in improved agreement between $E^{\rm int}_{\rm ISAPT}$  and DLPNO-CCSD(T)-based $E_{\rm ip}$ with
a mean deviation of 2.73 $\pm$ 2.49 kcal/mol (deviation of $\thickapprox 2\%$) for the 9 C$_x$L$_3$A$_y$ molecules.
The remaining discrepancies are presumably within the uncertainty of the model 
arising from averaging over mutation pathways (see FIG.\ref{fig01}b).

\begin{figure}[!hbpt]
    \centering
    \includegraphics[width=\linewidth]{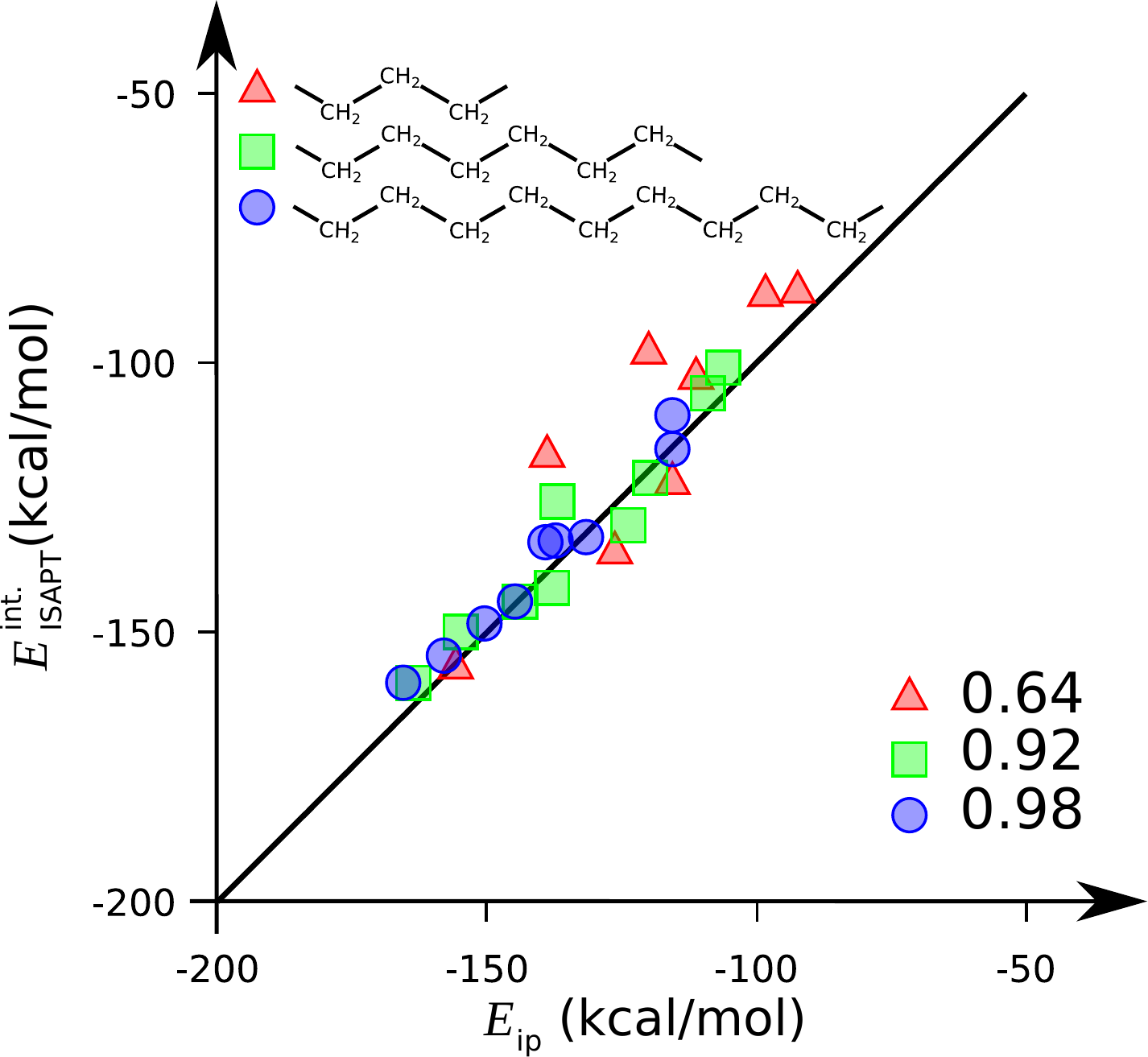}
    \caption{
    Scatterplot comparison of intramolecular ion-pair interaction energies, estimated with
    ISAPT, $E^{\rm int}$, and $E_{\rm ip}$, estimated using the thermodynamic
    cycle (Eq.~\ref{eq03}) based on $\omega$B97X-D3/def2-SVPD total energies. All values are in kcal/mol.
    Red triangles, green squares, and blue circles correspond to C$_x$L$_1$A$_y$,
    C$_x$L$_2$A$_y$, and C$_x$L$_3$A$_y$ systems, respectively. For each case,
    Pearson correlation coefficients are also provided. }
    \label{fig05}
\end{figure}
\begin{figure*}[!hbtp]
    \centering
    \includegraphics[width=\linewidth]{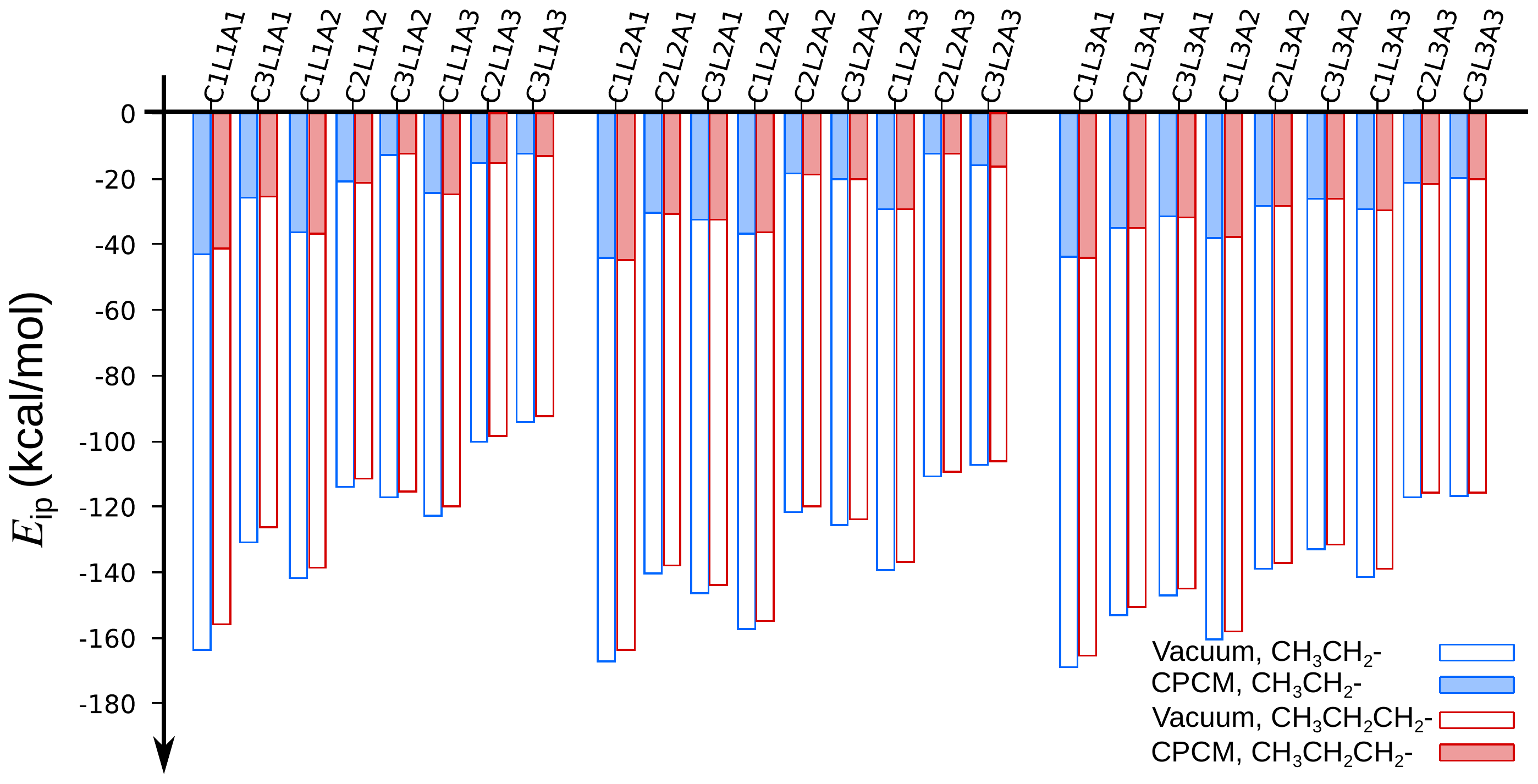}
    \caption{Effect of increasing the terminal group length, from ethyl to propyl,
    on $\omega$B97X-D3/def2-SVPD-based $E_{\rm ip}$ values in vacuum and with the CPCM
    implicit solvation model.}
    \label{fig06}
\end{figure*}

When replacing the DLPNO-coupled-cluster energies in Eq.~\ref{eq03} with DFT values, we see good agreement for basis sets with
a diffuse function: def2-SVPD and def2-TZVPD. For the L$_3$ systems with the longest terminal group explored,
the mean deviations of the $\omega$B97X-D3-DFT based estimations of $E_{\rm ip}$ from the ISAPT values are 3.25 and 2.97 for def2-SVPD and def2-TZVPD basis sets, respectively. Hence in further applications, we use the def2-SVPD basis set due to its
favorable cost-accuracy trade-off.

In FIG.~\ref{fig05}, we compare the absolute values of $E^{\rm int}_{\rm ISAPT}$ 
and $\omega$B97X-D3/def2-SVPD-based $E_{\rm ip}$ for 26 benchmark ion-pairs.
$E_{\rm ip}$ values were estimated using the $-$CH$_2$CH$_2$CH$_3$ terminal group.
Interaction energies computed using both methods lie in the range of -160 to -80 kcal/mol. Such
large values are expected in the vacuum phase as noted previously in the EDA calculations of model
ion-pair systems\cite{phipps2017intuitive}. 
As noted in Table.~\ref{tab:basis_benchmark_1}, with increasing linker length, the agreement improves as reflected by the Pearson correlation coefficients. The correlation is 0.98 when using a sufficiently long linker.
In general, the magnitude of
interaction energy is large when the polar groups are connected by a long and flexible linker resulting in a more compact structure.

The best agreement between F/ISAPT0 and the thermodynamic cycle is seen for the 9 C$_x$L$_3$A$_y$ systems with the X/Y=CH$_2$CH$_2$CH$_3$ terminal groups, in Table.~\ref{tab:DFT_benchmark_2}. Hence, we benchmark the performance of 10 DFT approximations only for these systems.
Several of the range-separated and long-range corrected hybrid-DFT methods show good agreement with the 
reference F/ISAPT0 values.
All methods (barring LC-PBE) differ amongst each other within 1 kcal/mol suggesting the 
uncertainty due to the choice of the DFT method to be smaller than the inherent uncertainty in the thermodynamic cycle. 
Since $\omega$B97X-D3 was used for geometry optimization,  
we continue with the same method in combination with the def2-SVPD basis set for the rest of the study.

Motivated by the good agreement with the DFT-based thermodynamic cycle and F/ISAPT0  gas-phase interaction energies, we now 
move on to see the effect of solvent on $E_{\rm ip}$. 
This is relevant because salt-bridging are mostly found in proteins, which in turn are often experimentally isolated in the aqueous phase.
Alas, the ISAPT-formalism has not been extended to solvent phase. However, there is
no limitation in applying the thermodynamic cycle in the solvent phase. 
The importance of accurately modeling the aqueous phase to study non-covalent interactions\cite{bootsma2018tuning,bootsma2019predicting}, reactions\cite{maldonado2021quantifying}, and ligand--metal interactions\cite{gentry2021computational} has been well established.
The magnitudes of $E_{\rm ip}$ 
is expected to decrease in the aqueous phase where the ion-pair interaction is screened. 
To this end, using the vacuum phase 
geometries, we performed CPCM total energy calculations for all 26 benchmark systems. 

FIG.~\ref{fig06} presents 
$E_{\rm ip}$ calculated at $\omega$B97X-D3/def2-SVPD and $\omega$B97X-D3(CPCM; water)/def2-SVPD levels. While
the vacuum phase values are $< -80$ kcal/mol, in the CPCM phase even the strongest interaction has an $E_{\rm ip}$ 
of about -40 kcal/mol. In general, the magnitude of vacuum phase $E_{\rm ip}$ decreases by a factor of 3.6--9.3 in the CPCM phase.
Further, the effect of CPCM on $E_{\rm ip}$ preserves the qualitative trends in ion-pair strengths as noted in the vacuum phase. In both 
media, we note 
the interactions due to monodentate polar moieties (ammonium or oxide) resulting in the strongest ion-pair interaction. 
We explored the ion-pair interactions with longer terminal groups ($-$CH$_2$CH$_3$ and $-$CH$_2$CH$_2$CH$_3$), and noted their influence on $E_{\rm ip}$ to be minimal in the CPCM phase.
FIG.~\ref{fig06} ascertains that our estimation of $E_{\rm ip}$ is efficient in the continuum solvation paradigm capturing commonly expected physical trends. This motivated us to investigate $E_{\rm ip}$ for some biologically relevant model systems
with CPCM and with microsolvation modeling to account for explicit solute-solvent interactions. 
\begin{figure}[!htbp]
    \centering
    \includegraphics[width=\linewidth]{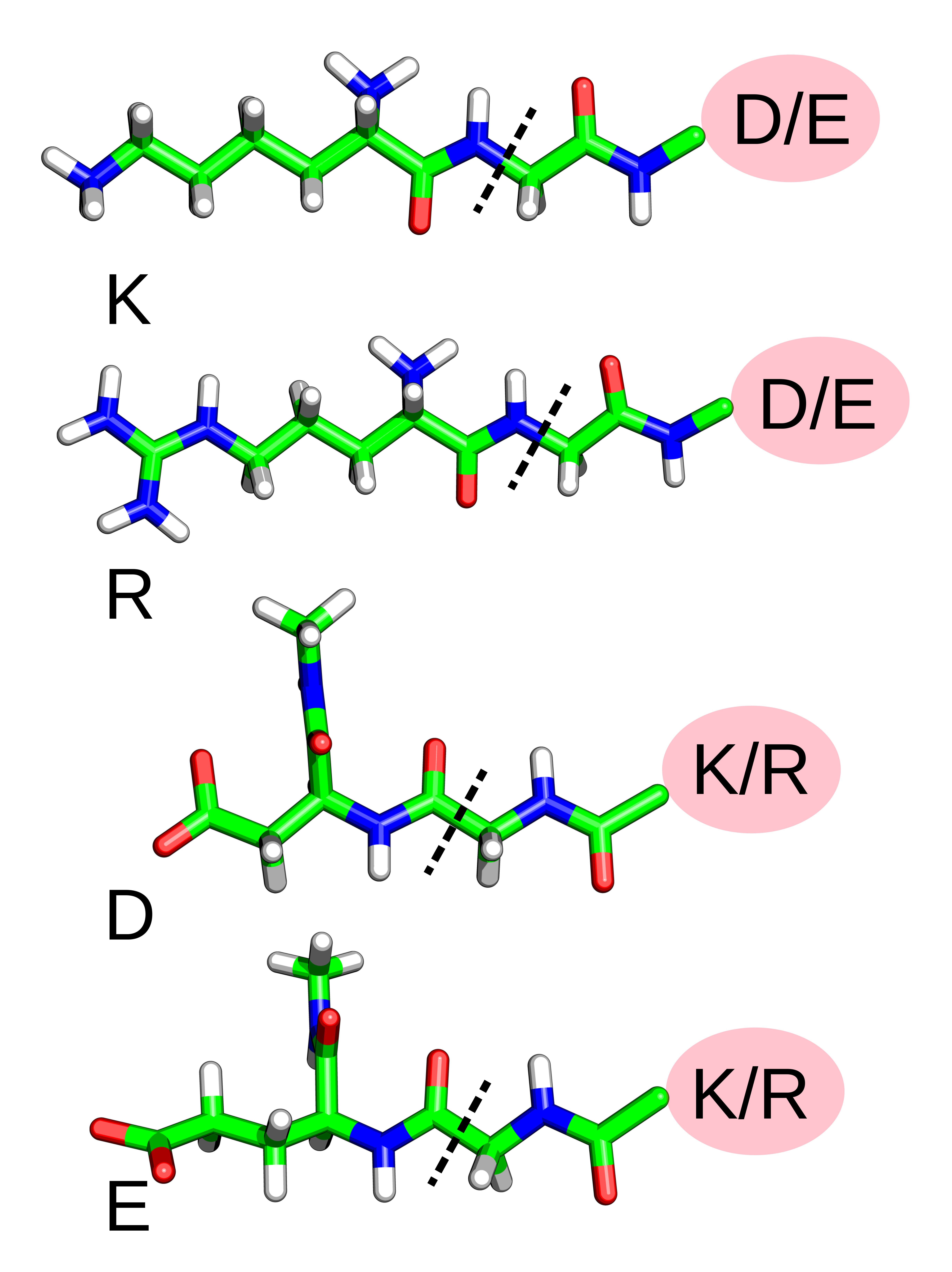}
    \caption{
    Definitions of terminal groups for charged amino acid residues:
    lysine (K), arginine (R), aspartic acid (D), and glutamic acid (E). For both N/C-terminal polar residues, all
    atoms until the C$_{\alpha}$ atom of the adjacent residue are included in the terminal groups by saturating
    the dangling bond with H. For the tripeptides, KGD, KGE, RGE, and RGD, the
    carboxylic acid group in the main chain is amidated by methyl amine.}
    \label{fig07}
\end{figure}

\subsection{Microsolvation effects on model tripeptides}

\begin{figure*}
    \centering
    \includegraphics[width=\linewidth]{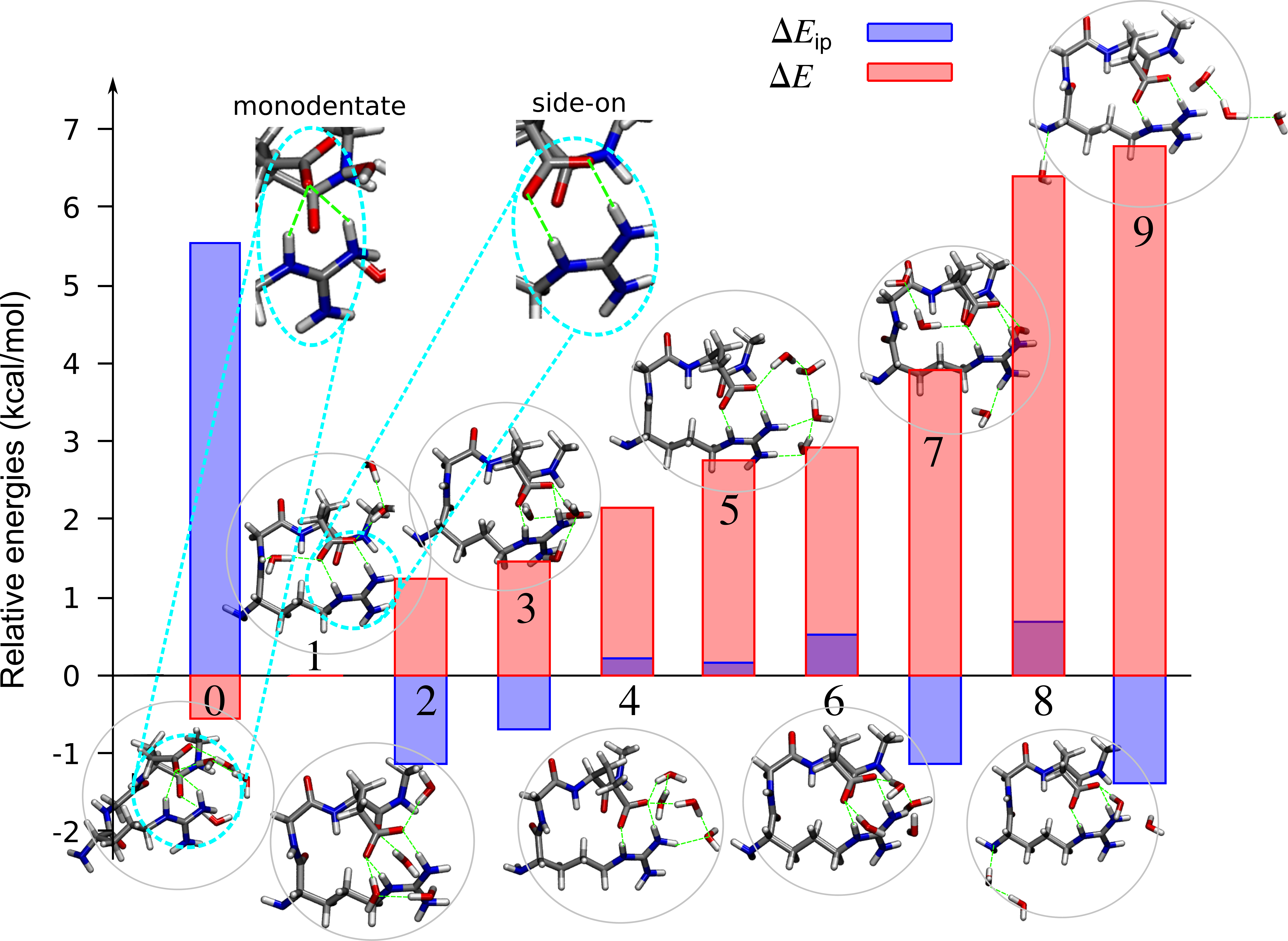}
    \caption{Relative energy, $\Delta E$, and relative ion-pair interaction energy, $\Delta E_{\rm ip}$, of 10 low-energy conformers of 
    microsolvated RGE (arginine-glycine-glutamic acid) estimated using
    thermodynamic cycles based on $\omega$B97X-D3(CPCM; water)/def2-SVPD total energies. All energies are reported by subtracting the values of the most stable conformer with a side-on guanidinium (arginine) - carboxylate (glutamic acid) salt-bridge. Minimum
    energy structures of all 10 conformers of RGE(H$_2$O)$_4$ are shown.
    Conformers are labelled from 0-9 (0 being the global minimum). 
    For clarity, structures have been arranged above and below the horizontal axis.
    The leftmost structure is with a monodentate salt-bridge 
    bonding where only one of
    the two O atoms of the carboxylate moiety
    binds to both NH atoms (see inset) while the structure right to it shows a bidentate interaction (see inset).
    } 
    \label{fig08}
\end{figure*}
Upon realizing the quenching effect of CPCM on $E_{\rm ip}$ in the 26 model systems we expand our study towards biologically relevant molecules with implicit and microsolvation modeling. 
The functionality of most macromolecules is strongly influenced by their interaction with
water molecules in the medium. For example, explicit solute-solvent interactions 
influence pharmacodynamic activities of ligand-protein complexes\cite{garcia2013hydration}.
It has been shown that quantitatively accurate modeling
of salt-bridges realistic an adequate description of the solvent phase\cite{pluhavrova2012peptide}. 
Since previous studies have shown that the zwitterionic form of amino acids is best described in the presence of water molecules\cite{schwaab2022zwitter,pluhavrova2012peptide}, we modeled the four 
tripeptides---lysine-glycine-aspartic acid (KGD), lysine-glycine-glutamic acid (KGE), arginine-glycine-aspartic acid (RGD), and arginine-glycine-glutamic acid (RGE)---with various degrees of microsolvation. 
Their C$-$terminals were amidated with methylamine to confine salt-bridging interactions to only between side-chains
and to prevent the participation of main chain carboxylic groups\cite{kumar2002close}.

\begin{figure*}[!htbp]
    \centering
    \includegraphics[width=\linewidth]{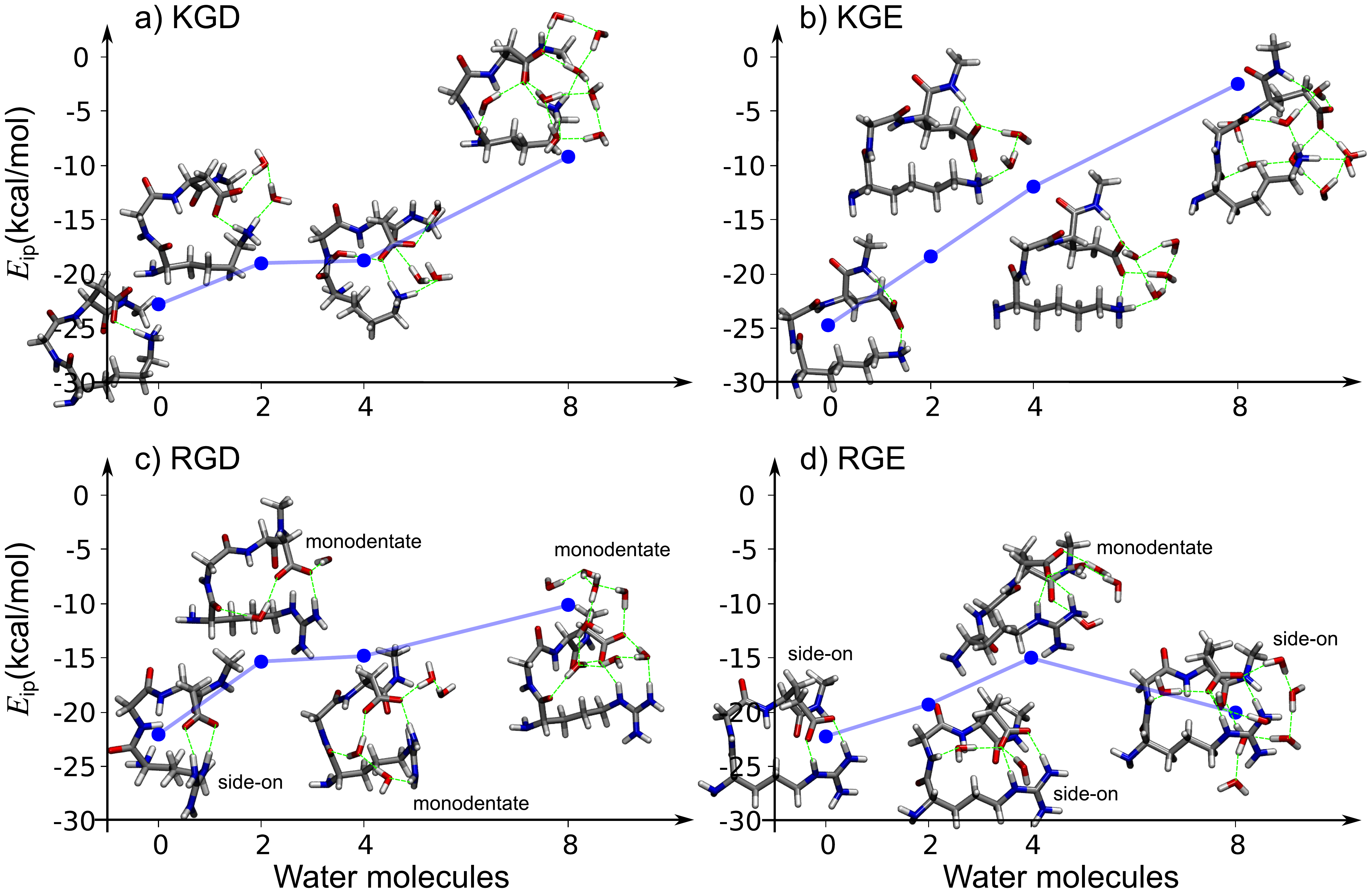}
    \caption{Effect of microsolvation on intramolecular ion-pair interaction energy in four tripeptides: KGD, KGE, RGD, and RGE.
    In all cases, 0-water corresponds to CPCM; water implicit solvation model. With 2, 4, and 8 explicit water molecules, 10 conformers were identified for which structures 
    and $E_{\rm ip}$ of the lowest energy conformer are shown. Guanidinium-carboxylate salt-bridges are classified as
    monodentate or side-on. Green lines denote ion-pair/H-bonding.}
    \label{fig09}
\end{figure*}
In FIG.~\ref{fig07}, we show the definitions of the polar terminal groups of for the four tripeptides studied here. 
We fragmented the positive amino acids by taking 
all atoms of the amino acid up to the C$_{\alpha}$ atom of the adjacent residue, {\it i.e.} glycine in this case. Hence, the NH part of the K-G and R-G amide bonds were included as a part of the cationic fragments, K and R. 
Similarly, the CO part of the DG and EG amide bonds were included as a part of the anionic fragments, D and E.
The dangling bonds were saturated by capping with an H atom. 
Hence, the terminal groups are now different for the two non-interacting components of {\it fr} (in FIG.~\ref{fig01}b) 
and they depend on the exact sequence of amino acids running from the N to C-terminals. 
This implies that the terminal groups for fragments of KGD and DGK will be different. 
The coordinates of the atoms in these fragments were kept frozen as in the compact structures and the mutation paths were followed as discussed before in Sec.\ref{Theory}.

In order to study the influence of water molecules on $E_{\rm ip}$, we inspected the 10 conformers of RGE in the presence of 4 water molecules in a background CPCM environment.
While the influence of conformers on intramolecular H-bonding interactions has been studied before\cite{karas2017trends}, here we wanted to study the conformer influence on intramolecular salt-bridging interactions.
We started with RGE relaxed in CPCM without any
explicit water molecule as the initial structure to which we added water molecules. 
Hence, when the initial structure showed a bidentate side-on\cite{huerta2015structure} interaction between guanidinium and carboxylate groups, it is surprising to observe in FIG.~\ref{fig08} a monodentate structure as the global minima.
Previous studies have noted a better stabilization of solvent-separated ion-pair in lysine-glutamate dipeptide with an increasing number of explicit water molecules\cite{pluhavrova2012peptide}. 
The remaining 9 complexes showed side-on interactions.
We calculated the relative $E_{\rm ip}$ values with reference to the most stable conformer with a side-on interaction. 

\begin{figure*}[!htbp]
    \centering
    \includegraphics[width=\linewidth]{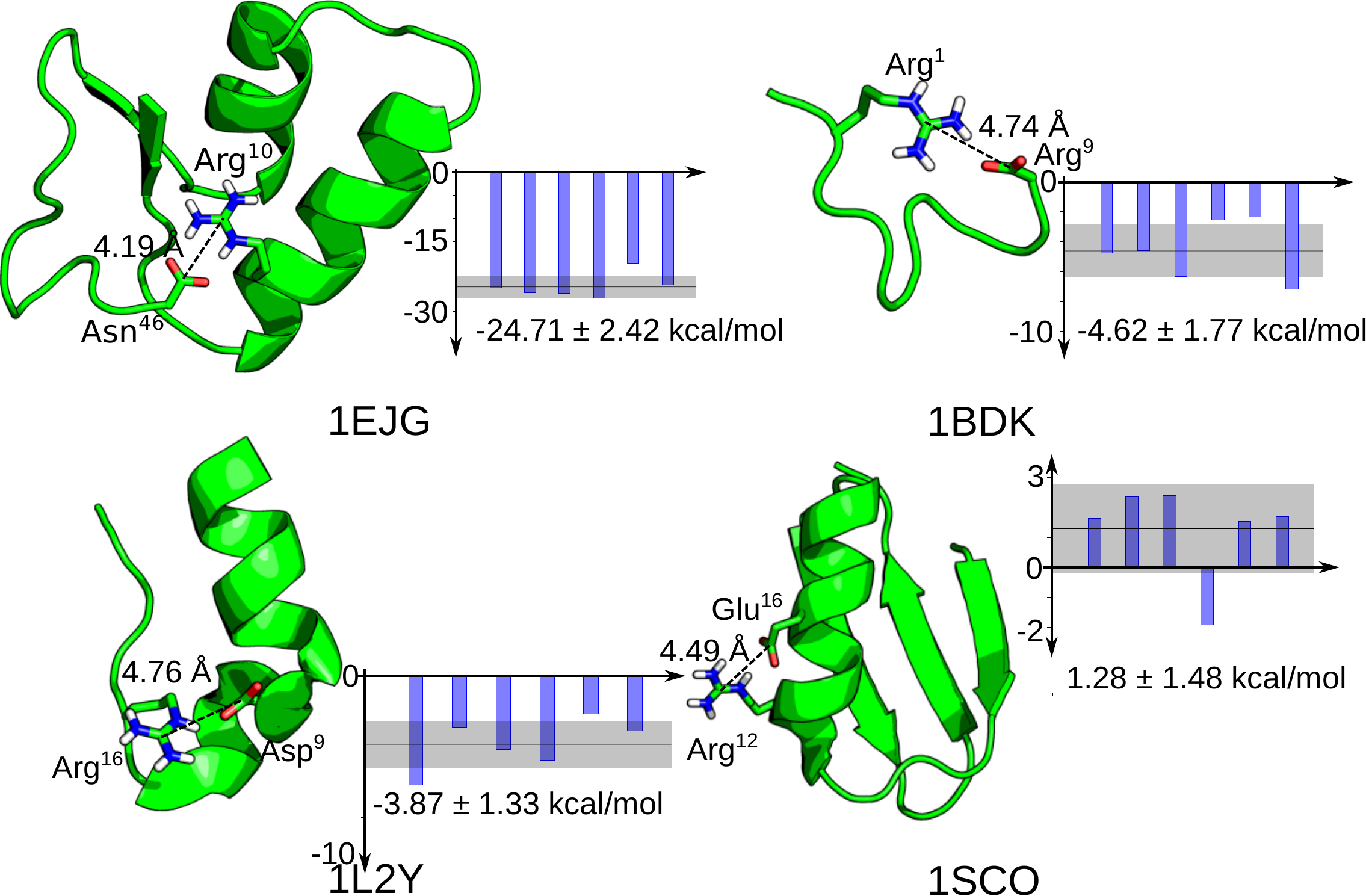}
    \caption{Strength of guanidinium--carboxylate salt-bridge in the peptides
    1EJG, 1BDK, 1L2Y and 1SCO estimated using 
    a thermodynamic cycle with 
    $\omega$B97X-D3(CPCM; water)/def2-SVPD total energies. 
    The centroid--centroid distance between the interacting 
    moieties are also given.
    In each case, 6 possible alchemical mutations result 
    in different values of $E_{\rm ip}$
    shown in histograms. The net salt-bridge interaction energy corresponds to the average marked by horizontal line and the standard deviation is denoted by an envelope. 
    }
    \label{fig10}
\end{figure*}

While for all side-on conformers, the $\Delta E_{\rm ip}$ is very similar, for the monodentate complex it is found to be de-stabilizing.
Inspecting this structure revealed that this is due to an unfavorable orientation that diminishes the salt-bridging interaction.
The net stability of the global minimum is due to solvation energy dominating over the salt-bridging contribution.
It may be realized that for a highly stable complex, the water molecules and the polar groups must involve in maximal bonding (including
H$-$bond network among the water molecules).
The role of solvation in stabilizing a system decreases when the
H$-$bonds are formed with weakly polar groups or when the water network is disconnected.
In FIG.~\ref{fig08}, we note this effect on the conformers on the right side, where the complexes 
destabilize when all solvent molecules do not form extended networks with the polar groups.
Therefore, it may be concluded that, while favorable orientations with water molecules may benefit the overall
stability of the complex, it comes at the cost of quenched salt-bridging interactions. 

Varying the number of water molecules in the microsolvation of tripeptide will provide 
insights into the effect of solvation in the bulk water scenario.
In FIG.~\ref{fig09}, we plotted $E_{\rm ip}$ for all 4 tripeptides in the presence of 0, 2, 4, and 8 water molecules. 
The zero-water case corresponds to the CPCM limit with one minimum energy structure. 
For the case of 2, 4, and 8 water complexes, we sampled 10 conformers starting with different arrangements of water molecules,
of which we report only the global minimum's $E_{\rm ip}$ regardless of the bonding denticity. 
Thus, for all 4 peptides, increasing the number of water molecules mostly results in quenched salt-bridging interactions. 
This is a physically expected trend that has been captured by the model. However, the degree to which the $E_{\rm ip}$ changes is different across the peptides. This can be attributed to the relatively small conformer space sampled and 
due to the finite degree of hydration.

A closer look at the structures sheds more light on the magnitude of the interaction 
energies reported in FIG.~\ref{fig09}. For KGD, with an increasing number of water molecules, 
the interactions between the polar fragments diminish. 
While this trend is seemingly similar in KGE, in the case of 8 water molecules KGE adopts a solvent separated ion-pair form, while
that of KGD is solvent-shared with a stronger ion-pair interaction. 
For RGD, all solvated complexes prefer the monodentate arrangement. Hence with increasing degree of hydration,
$E_{\rm ip}$ of RGD approaches a limit of $> -10$ kcal/mol. In the case of RGE, 0, 2, and 8 water complexes form a side-on
bidentate arrangement, while with 4 water molecules, the most stable conformer has a monodentate arrangement with an elevated
$E_{\rm ip}$. It may be noted that the linker length is longer in RGE compared to RGD favoring a more stable
side-on salt-bridging. Hence, by comparing the bonding pattern of the minima we can qualitatively interpret $E_{\rm ip}$ values estimated using the thermodynamic cycle.

\subsection{Application to large peptides}
We collected the NMR resolved structures of crambin (1EJG)\cite{mandal2012design}, bradykinin (1BDK)\cite{sejbal1996nmr}, tryptophan zipper I (1L2Y)\cite{neidigh2002designing},
and a scorpion toxin OSK1 (1SCO)\cite{jaravine1997three}. 
Crambin (1EJG) is a neutral plant protein with 46 amino-acid residues and 642 atoms stabilized by the salt-bridging interaction between a 
carboxylate group on the main chain of Asn46 and 
a guanidinium group of Arg10. 
1BDK is 11 residues long with 187 atoms with a net charge of +3. 
1L2Y has a salt-bridging interaction between Arg16 and Asp9, containing 304 atoms and a net charge of +1.
1SCO is a scorpion toxin with an overall charge of +8 and 595 atoms, presenting a 
$i$-to-$i$+4 salt-bridge between Arg12 and Glu16. These 4 peptides provide an opportunity to 
understand the significant role orientation plays in deciding the strength of a salt-bridge.
Their structures are on display in FIG.~\ref{fig10}. In these proteins,
centroid-centroid distances of ion-pairs are $<$ 5.0\AA. 
1EJG shows a side-on salt-bridge (Asn46-Arg10) with a short centroid-centroid distance leading to a significantly large magnitude of $E_{\rm ip}$, -24.7 kcal/mol. 
1BDK is stabilized in a monodentate fashion with a relatively weaker interaction energy than 1EJG. This is followed by 1L2Y presenting an even more diminished interaction. Finally, in 1SCO we note a small positive $E_{\rm ip}$. 

A closer inspection of the individual $E_{\rm ip}$ values
 obtained through various mutation paths provide us with further insight. As noted before in FIG.~\ref{fig01}c, a salt-bridge with a symmetric bonding pattern should show a smaller variance across mutation paths. Hence, for 1EJG, where the salt-bridge is symmetric, all $E_{\rm ip}$ obtained are of similar magnitude with a small standard deviation. However, for monodentate arrangements in 1BDK and 1L2Y, the symmetry is lost and we note a larger variation in the $E_{\rm ip}$. For 1SCO, while most mutation paths yield small positive values, there is one path that yields a small negative value. Hence, the salt-bridging in this system is essentially
 non-interacting, if not mildly unfavorable. 
 This could be speculated to be due to the unfavorable orientation of the interacting moieties. 
 Despite the relatively short centroid-centroid distance, the polar groups are oriented away from each other, resulting in a net repulsion, unlike all the other cases explored here.
 The small positive values arise when we explore mutation paths that involve these unfavorably oriented atoms.
 This example demonstrates how a salt-bridging with a reasonably small centroid-centroid distance can be overall irrelevant.
 
\section{Conclusions}
In this study, we present a thermodynamic cycle to quantify intramolecular ion-pair
interactions between two polar moieties of a molecule, without complex bifurcating interactions.
We benchmark the efficiency of this approach by comparing intramolecular interaction energies obtained for a set of 26 zwitterions stabilized by ion-pair interactions against predictions from ISAPT. We found the deviations between the two approaches to be minimal for molecules 
containing long linkers and when the terminal groups are adequately capped. Going from the vacuum phase to the aqueous 
phase, the ion-pair interactions are quenched as expected. 

We modeled microsolvation of four biologically relevant tripeptides and found 
increasing degree of solvation to quench their $E_{\rm ip}$. 
In the presence of solvent molecules, the weakest interaction was found for a solvent-separated salt-bridge where the polar groups are
surrounded by a cage of water molecules. The salt-bridge interaction is stronger for conformers
with a bidentate side-on bonding. On the other hand, in conformers with a monodentate bonding, the salt-bridge is more exposed to solvent molecules resulting in better hydration of the biomolecule. 
In the aqueous environment both hydration and salt-bridging are
competing stabilizing factors, the latter favoring a folded structure. 
We applied the thermodynamic cycle on experimental structures of 4 large proteins
and found the salt-bridging interaction to be strongest for 1EJG with a side-on bonding, 
while 1BDK and 1L2Y with a monodentate bonding showed weak interactions. 
In 1SCO, even though the distance between polar groups suggests a possible
salt-bridge bonding, due to the unfavorable orientation of the corresponding moieties, its $E_{\rm ip}$ turned out to be insignificant. 
The favorable computational overhead of the proposed method promises explorations of
salt-bridging interaction in other biological systems in the solvent phase.
Since the present model can only account for simple, non-bifurcated salt-bridges, future endeavors may concentrate on quantifying interactions for complex salt-bridges.

\begin{acknowledgements}
We acknowledge support of the Department of Atomic Energy, Government
of India, under Project Identification No.~RTI~4007. 
All calculations have been performed using the Helios computer cluster, 
which is an integral part of the MolDis Big Data facility, TIFR Hyderabad \href{http://moldis.tifrh.res.in}{(http://moldis.tifrh.res.in)}.
\end{acknowledgements}

\section{Author Declarations}
\subsection{Conflicts of Interest}
The authors have no conflicts of interest to disclose.
\subsection{Data Availability}
The data that support the findings of this study are openly available in Github, \Refs{SI_IntraIonPair}.

\section*{References}
\bibliography{salt_bridge.bib}
\end{document}